\documentclass{JHEP3} 

\usepackage{epsfig,multicol}
\usepackage{amsmath, amssymb}

\usepackage{concmath, palatino}

\newcommand\fverb{\setbox\pippobox=\hbox\bgroup\verb}
\newcommand\fverbdo{\egroup\medskip\noindent%
			\fbox{\unhbox\pippobox}\ }
\newcommand\fverbit{\egroup\item[\fbox{\unhbox\pippobox}]}
\newbox\pippobox

\newcommand{\be}{\begin{equation}} 
\newcommand{\ee}{\end{equation}}

\newcommand{\ba}{\begin{eqnarray}}
\newcommand{\ea}{\end{eqnarray}}

\newcommand{\la}{\longrightarrow}

\newcommand{\ads}{AdS_5\times S^5}

\newcommand{\refeq}[1]{Eq.~(\ref{#1})}

\title{Universality of three gaugino anomalous dimensions in ${\cal N}=4$ SYM}

\author{Matteo Beccaria\\
  Dipartimento di Fisica, Universita' di Lecce,
  Via Arnesano, 73100 Lecce\\
  INFN, Sezione di Lecce\\
  E-mail: \email{matteo.beccaria@le.infn.it}}

\preprint{}

\abstract{
We study maximal helicity three gaugino operators in ${\cal N}=4$ Super Yang-Mills theory. We show that the lowest anomalous dimension 
of scaling operators with generic finite spin can be expressed in terms of the universal anomalous dimension appearing at twist-2. This statement is 
rigourously 
proved at three loops. The reason for this universality between sectors 
with different twist is the hidden $\mathfrak{psu}(1|1)$ invariance
of the $\mathfrak{su}(2|1)$ subsector of the theory.
}

\begin{document} 

\section{Introduction}
\label{sec:Intro}

The renormalization flow of certain QCD closed sectors of composite operators is perturbatively integrable in the planar 
limit~\cite{Belitsky:2004cz}. This intriguing feature can be studied in QCD-like theories with supersymmetry
where integrability can be understood in the spirit of AdS/CFT duality~\cite{Maldacena:1997re} in terms of the integrability properties
of the dual superstring theory on $AdS_5\times S^5$~\cite{Bena:2003wd}.
In particular, one can consider the maximally supersymmetric ${\cal N}=4$ super Yang-Mills theory 
which is UV finite and superconformally invariant at the quantum level. There, asymptotic all-loop Bethe Ansatz equations
are available for the full set of $\mathfrak{psu}(2,2|4)$ operators~\cite{Beisert:2005fw}.

\medskip
Particularly interesting is the bosonic non compact $\mathfrak{sl}(2)$ sector. The three loop Bethe Ansatz equations were derived in~\cite{Staudacher:2004tk}.
Their prediction have been cofirmed by independent field theoretical checks at two loops
in~\cite{Eden:2005bt}. The two-loop dilatation operator has also been constructed algebraically in the $\mathfrak{su}(1,1|2)\supset \mathfrak{sl}(2)$
sector~\cite{Zwiebel:2005er} and explicitly in the $\mathfrak{sl}(2)$ sector~\cite{Eden:2006rx}. 

\medskip
With a major breakthrough, Kotikov, Lipatov, Onishchenko and Velizhanin conjectured a three loop prediction for the anomalous dimension of  ${\cal N}=4$ 
twist-2 superconformal operators at generic spin in~\cite{Kotikov:2004er} (see also~\cite{klov}). Their prediction is based the so-called {\em maximum transcendentality principle}. 
Recently, the principle has been extended and allowed the calculation of the four loop anomalous dimension of twist-3 operators in the $\mathfrak{sl}(2)$
sector~\cite{Beccaria:2007cn,Kotikov:2007cy}.

\medskip
A different approach to integrable systems, related to the Bethe Ansatz, is based on the Baxter 
$Q$-operator~\cite{Bax72}. Using this method, the two loop dilatation operator in ${\cal N}=2, 4$ SYM
for Wilson operators with scalars and derivatives (the $\mathfrak{sl}(2)$ sector) as well as the two loop Baxter 
operator have been computed in~\cite{Belitsky:2006av}. The all-loop asymptotic generalization of the Baxter equation
appeared in~\cite{Belitsky:2006wg}. All-loop extensions to the $\mathfrak{sl}(2|1)$ sector are described in ~\cite{Belitsky:2006cp,Belitsky:2006cp}.

\medskip
These results strongly support integrability as a quite efficient computing tool for the calculation of multi-loop anomalous dimensions. 
However, it seems that the $\mathfrak{psu}(2,2|4)$ symmetry still has to be fully exploited. This is clear in the deep discussion of~\cite{Beisert:2005fw}
about degeneracies in the spectum of anomalous dimensions. The Bethe Ansatz equations have remarkable structural properties related to supersymmetry.
Degeneracies appear relating sectors built with composite operators with a different number of elementary fields. 

\medskip
This is far beyond what is well known in twist-2. There, all conformal operators fall in a single supermultiplet~\cite{Belitsky:2003sh,Belitsky:2005gr,Beisert:2002tn} 
and the anomalous dimension of different channels are related by supersymmetry
and can be expressed in terms of a single universal function $\gamma_{\rm univ}$. The methods of~\cite{Beisert:2005fw} suggest instead
that one can expect hidden relations between operators with different twists.

\medskip
In this paper, we present a nice example of this mechanism. We consider twist-3 composite operators built with three gauginos. These 
operators have been studied at two loops in ${\cal N}=1, 2, 4$ SYM by direct computation of 
the dilatation operator in~\cite{Belitsky:2005bu}. We provide a simple exact formula for the one-loop lowest anomalous dimension. It matches the 
universal function $\gamma_{\rm univ}$ with suitable shifted argument. 

\medskip
We remark that this degeneracy linking operators with different spin has already appeared in the literature. It has been discussed 
by Korchemsky and coworkers~\cite{Korchemsky:1994um,Derkachov:2002wz}
in the analysis of the spectrum of compound states of reggeized gluons in planar QCD. There, one is lead to study the ground states of a 
noncompact Heisenberg $\mathfrak{sl}(2, \mathbf{C})$ spin magnet. The degeneracies discovered in those work cover the one studied in this paper, 
although only  at the one-loop level. 

\medskip
As a further step, we prove that this universal relation is valid at three loops because of supersymmetry. We prove this fact by an explicit analysis of the 
relevant Bethe Ansatz equations. This is feasible at three loops. Due to the symmetry related reason of this universality, we feel that is should be 
possible to prove it at all orders in terms of supermultiplet rearrangements. We also give a two-loop proof based on the Baxter formalism, 
as an extension of the results of~\cite{Korchemsky:1994um,Derkachov:2002wz}. 

\medskip
The detailed plan of the paper is as follows. Sec.~(\ref{sec:one}) is devoted to a complete analysis at one-loop.
Sec.~(\ref{sec:two}) collects some known two-loop results and shows that they are in agreement with the one-loop universality.
Sec.~(\ref{sec:three}) proves universality at three loops by analyzing the Bethe Ansatz equations. Finally, Sec.~(\ref{sec:baxter})
is devoted to a similar two-loop proof at the level of the Baxter equation.

\section{One-loop anomalous dimension of quasi-partonic operators}
\label{sec:one}

We adopt the notation of~\cite{Belitsky:2006en} and  consider the following class of single-trace conformal Wilson operators 
\be
{\cal O}_{s, L}(0) = \sum_{n_1+\cdots n_L=s} a_{n_1,\dots n_L}\,\mbox{Tr}\left\{D_+^{n_1} X(0)\cdots D_+^{n_L} X(0)\right\},\ \ n_i\in\mathbb{N},
\ee
where $X(0)$ is a physical component of quantum fields with definite helicity in the underlying gauge theory (scalar, fermion or gauge field),
and $D_+$ is a light-cone projected covariant derivative. The coefficients $\{a_\mathbf{n}\}$ determine eigenoperators of the dilatation operator.
The total Lorentz spin is $s=n_1 + \cdots n_L$. The twist $L$ is, as usual, the classical dimension minus the Lorentz spin.

\medskip
At one-loop, it is well-known that the anomalous dimensions of the above operators are in 1-1 correspondence with the spectrum of a 
noncompact $\mathfrak{sl}(2)$ spin chain with $L$ sites. The elementary spin of the chain is related to the conformal spin $\eta$ of $X$
which is defined as $\eta = \frac{1}{2}, 1, \frac{3}{2}$ when $X$ is a scalar, gaugino, or gauge field respectively.

\medskip
The one-loop ground state energy, associated with the lowest anomalous dimension, can be found easily by the Baxter approach~\cite{Bax72}.
The Baxter function is a polynomial $Q(u)$  satisfying the second-order finite-difference equation
\be\label{Baxter-eq}
(u + i\,\eta)^L\, Q (u + i) + (u - i\,\eta)^L\, Q(u - i) = t_L(u)\, Q(u).
\ee
Here $t_L(u)$ is a polynomial in $u$ of degree $L$ with coefficients given by conserved charges
\be\label{t_L}
t_L(u) = 2\, u^L + q_2\, u^{L-2} + \ldots + q_L
\ee
The lowest integral of motion $q_2$ is related to the total spin of the $\mathfrak{sl}(2)$
chain, $s+L\eta$,
\be\label{q2}
q_2 = -(s+L\,\eta) (s+L\,\eta - 1) + L\, \eta\,(\eta-1),
\ee
with $s=0,1,\ldots$.

\medskip
In what follows we shall refer to \refeq{Baxter-eq} as the Baxter equation. The degree of $Q(u)$ is equal to the total spin $s$. Up to 
an irrelevant normalization, one can write
\be\label{Q-polynom}
Q(u) = \prod_{k=1}^s (u-\lambda_k)\,.
\ee
If one replaces this expression into \refeq{Baxter-eq}, the roots $\lambda_1,\ldots,\lambda_s$ are found to obey the Bethe equations
\be\label{Bethe-roots}
\left(\frac{\lambda_k+i\,\eta}{\lambda_k-i\,\eta}\right)^L=\mathop{\prod_{j=1}^s}_{j\neq k}
\frac{\lambda_k-\lambda_j-i}{\lambda_k-\lambda_j+i}\,.
\ee
Solving the Baxter equation \refeq{Baxter-eq} supplemented by \refeq{Q-polynom} one
obtains quantized values of the charges $q_3,\ldots,q_L$ and evaluates the
corresponding energy and quasimomentum as 
\be\label{Energy-Baxter}
\varepsilon = i\left(\ln Q(i\,\eta)\right)'-i\left(\ln Q(-i\,\eta)\right)'\,,\qquad e^{i\theta} =
\frac{Q(i\,\eta)}{Q(-i\,\eta)}\,.
\ee
The cyclic symmetry of the single-trace operators imposes an
additional selection rule for the eigenstates of the spin magnet,
$e^{i\theta}=1$. Equations \refeq{Energy-Baxter} allows to
calculate the energy of the spin chain and, then, obtain the one-loop anomalous
dimension of Wilson operators  using
\be
\Delta \gamma(s) = g^2\,\varepsilon(s) + {\cal O}(g^4),
\ee
where $g^2 = g_{\rm YM}^2\,N_c/(8\,\pi^2)$ is the scaled 't Hooft coupling, fixed in the planar $N_c\to\infty$ limit.
In the above expressions, $\Delta\gamma(s) = \gamma(s)-\gamma(0)$ is the subtracted anomalous dimension defined in order to vanish at $s=0$.

\subsection{Twist-2}

Solving the Baxter equation at twist-2 in the three sectors $\eta = 1/2, 1, 3/2$, i.e. for the scalar, gaugino and vector channels denoted by 
the symbols $\varphi, \lambda, A$, one immediately recovers the known formulae
\ba
\Delta\gamma_{L=2}^\varphi(s) &=& 4\,S_1(s), \nonumber \\
\Delta\gamma_{L=2}^\lambda(s) &=& 4\,S_1(s+1)-4, \\
\Delta\gamma_{L=2}^A(s) &=& 4\,S_1(s+2)-6. \nonumber
\ea
Our notation for the (nested) harmonic sums is
\be
S_a(N) = \sum_{n=1}^N\frac{(\mbox{sign}\,a)^n}{n^a}, \qquad S_{a_1, a_2, \dots}(N) = \sum_{n=1}^N\frac{(\mbox{sign}\,a_1)^n}{n^{a_1}}\,S_{a_2, \dots}(n).
\ee
Alternative expressions with the $\psi$ functions are a little nicer and read
\ba
\Delta\gamma_{L=2}^\varphi(s) &=& 4\,(\psi(s+1)-\psi(1)), \nonumber \\
\Delta\gamma_{L=2}^\lambda(s) &=& 4\,(\psi(s+2)-\psi(2)), \\
\Delta\gamma_{L=2}^A(s) &=& 4\,(\psi(s+3)-\psi(3)). \nonumber
\ea
Notice that $\Delta\gamma\equiv\gamma$ in the scalar channel. 
These results express the well-known fact that all twist-2 quasipartonic operators are in the same SUSY multiplet
and their anomalous dimension is expressed by a universal function with shifted arguments
\ba
\gamma_{L=2}^\varphi(s) &=& \gamma_{\rm univ}(s), \nonumber \\
\gamma_{L=2}^\lambda(s) &=& \gamma_{\rm univ}(s+1), \\
\gamma_{L=2}^A(s) &=& \gamma_{\rm univ}(s+2). \nonumber
\ea
The universal function $\gamma_{\rm univ}(s)$ is known at three loops and reads
\be
\gamma_{\rm univ}(s) = \sum_{n\ge 1}\gamma_{\rm univ}^{(n)}(s)\,g^{2\,n} ,
\ee
where the three loop coefficients are~\cite{Kotikov:2004er}
\ba
\label{eq:univ}
\gamma_{\rm univ}^{(1)}(s) &=& 4\, S_1\, ,  \\
\gamma_{\rm univ}^{(2)}(s) &=&-4\,\Big( S_{3} + S_{-3}  -
2\,S_{-2,1} + 2\,S_1\,\big(S_{2} + S_{-2}\big) \Big)\, ,  
\nonumber \\
\gamma_{\rm univ}^{(3)}(s) &=& -8 \Big( 2\,S_{-3}\,S_2 -S_5 -
2\,S_{-2}\,S_3 - 3\,S_{-5}  +24\,S_{-2,1,1,1}\nonumber\\
&&~~~~~~+ 
6\,\big(S_{-4,1} + S_{-3,2} + S_{-2,3}\big)
- 12\,\big(S_{-3,1,1} + S_{-2,1,2} + S_{-2,2,1}\big)\nonumber \\
&&~~~~~~-
\big(S_2 + 2\,S_1^2\big) 
\big( 3 \,S_{-3} + S_3 - 2\, S_{-2,1}\big)
- S_1\,\big(8\,S_{-4} + S_{-2}^2\nonumber \\
&&~~~~~~+ 
4\,S_2\,S_{-2} +
2\,S_2^2 + 3\,S_4 - 12\, S_{-3,1} - 10\, S_{-2,2} 
+ 16\, S_{-2,1,1}\big)
\Big)\, , \nonumber
\ea
with all harmonic sums evaluated at argument $s$.

\subsection{Twist-3}

The same exercise at twist-3 gives
\ba
\Delta\gamma_{L=3}^\varphi(s) &=& 4\,S_1\left(\frac{s}{2}\right), \nonumber \\
\Delta\gamma_{L=3}^\lambda(s) &=& 4\,S_1(s+2)-6, \\
\Delta\gamma_{L=3}^A(s) &=& 4\,S_1\left(\frac{s}{2}+1\right)-5+\frac{4}{s+4}.\nonumber
\ea
Again, alternative expressions with the $\psi$ function are 
\ba
\Delta\gamma_{L=3}^\varphi(s) &=& 4\,\left[\psi\left(\frac{s}{2}+1\right)-\psi(1)\right], \nonumber \\
\Delta\gamma_{L=3}^\lambda(s) &=& 4\,(\psi(s+3)-\psi(3)), \\
\Delta\gamma_{L=3}^A(s) &=& 4\,\left[\psi\left(\frac{s}{2}+2\right)-\psi(1)\right]-5+\frac{4}{s+4}. \nonumber
\ea
The scalar channel is very well-known by now. Indeed, the four-loop expression of $\Delta\gamma_{L=3}^\varphi(s)$ has been 
recently computed in ~\cite{Beccaria:2007cn,Kotikov:2007cy}.

\medskip
The 3-gaugino scaling operator has an anomalous dimension which strongly reminds the twist-2 supermultiplet. 
This is a non-trivial effect of supersymmetry since it relates composite operators with a different number of fields.
As we mentioned in the Introduction, this one-loop degeneracy has an old story and has
been first studied in~\cite{Korchemsky:1994um,Derkachov:2002wz}.

\medskip 
The $L=3$ operator built with vector fields has an anomalous dimension which is not related to the other channels in any obvious way.
Also, it contains a peculiar rational contribution. Actually, this  expression is not
totally surprising. Three-gluon operators are studied in QCD in~\cite{Belitsky:1999bf}. The dilatation operator
has an integrable piece ${\cal H}_0$ plus a perturbation. The lowest eigenvalue of the integrable piece has 
eigenvalues given by Eq.~(82) of~\cite{Belitsky:1999bf}:
\ba
\varepsilon &=& 2\,\psi\left(\frac{s}{2}+3\right)+2\,\psi\left(\frac{s}{2}+2\right)-4\,\psi(1)+4 = \\
&=& 2\,S_1\left(\frac{s}{2}+2\right)+2\,S_1\left(\frac{s}{2}+1\right)+4 = \\
&=& 4\,S_1\left(\frac{s}{2}+1\right)+\frac{4}{s+4}+4.
\ea
Apart from the constant, this is the same $s$ dependent combination as in $\Delta\gamma_{L=3}^A$.

\medskip
Given these interesting one-loop results, one would like to show that the one-loop relation between the twist-3 gaugino channel and 
the universal twist-2 anomalous dimension is not an accident. Before proving it, let us illustrate some available and recent two loop results
that indeed support this conjecture.

\section{Additional evidence for universality at two-loops}
\label{sec:two}

Three gaugino operators have been studied at two loops in ${\cal N}=1, 2, 4$ SYM by direct computation of 
the dilatation operator in~\cite{Belitsky:2005bu}. For even spin $s$, the lowest anomalous dimension is that of an unpaired state
with zero quasimomentum. The $\overline{DR}$ anomalous dimension for $s=4, 6$ is reported as 
\ba
\Delta\gamma_{L=3}^\lambda(s=4) &=& \frac{19}{5}\,g^2 + \left(\frac{15581}{2250}-\frac{19}{5}\,{\cal N}\right) \,g^4 + \cdots, \\
\Delta\gamma_{L=3}^\lambda(s=6) &=& \frac{341}{70}\,g^2 + \left(\frac{55402939}{6174000}-\frac{341}{70}\,{\cal N}\right) \,g^4 + \cdots .
\ea
In this channel, we have, at two loops
\be
\gamma_{L=3}^\lambda(s) = \Delta\gamma_{L=3}^\lambda(s) + \Gamma^\lambda,
\ee
where the anomalous dimension of the three-gaugino operator without derivative, hence $s=0$, is given by 
\be
\Gamma^\lambda = 6\,g^2-12\,g^4 + \cdots.
\ee
Adding it to $\Delta\gamma$ and replacing ${\cal N}=4$, we get
\ba
\gamma_{L=3}^\lambda(s=4) &=& \frac{49}{5}\,g^2-\frac{45619}{2250}\,g^4 + \cdots, \\
\gamma_{L=3}^\lambda(s=6) &=& \frac{761}{70}\,g^2-\frac{138989861}{6174000}\,g^4 + \cdots.
\ea
Comparing with the one, two-loop expressions of $\gamma_{L=2}^\varphi$ we see that we can write in both cases
\be
\label{eq:conj}
\gamma_{L=3}^\lambda(s) = \gamma_{\rm univ}(s+2),\qquad s\in 2\mathbb{N}.
\ee
It is tempting to conjecture that this relation is actually valid at all orders and for any even spin $s$.
In the next section we shall provide a very simple and explicit proof that the conjecture holds true at least at the three loop level. The main tool will be the 
set of Bethe Ansatz equations in the $\mathfrak{sl}(2|1)$ subsector of the ${\cal N}=4$ theory.

\section{Proof of universality at three loops}
\label{sec:three}

Our proof builds on the results of~\cite{Belitsky:2007zp,Belitsky:2006cp}, whose notation we follow.
The $\mathfrak{sl}(2|1)$ sector of light-cone ${\cal N}=4$ SYM is a convenient truncation suitable for the proof of universality of $\gamma_{L=3}^\lambda$.
The elementary fields are a complex scalar $X(z)$ and single-flavour gaugino $\psi(z)$ with an arbitrary number of light-cone projected covariant derivatives.
The pair $(X, \psi)$ fills a chiral ${\cal N}=1$ multiplet
\be
\Phi(Z) = i\,X(z)+\theta\,\psi(z),\qquad Z=(z, \theta).
\ee
The composite fields in the planar limit are single-traces operators of the form 
\be
{\cal O}(Z_1, \dots, Z_L) = \mbox{Tr}\left\{\prod_{i=1}^L\Phi(Z_i)\right\},
\ee
and can be expanded in components to give scaling fields of the form 
\be
{\cal O}_{s, L}(0) = \sum_{n_1+\cdots n_L=s} a_{n_1,\dots n_L}\,\mbox{Tr}\left\{D_+^{n_1} \Omega_1(0)\cdots D_+^{n_L} \Omega_L(0)\right\},\ \ n_i\in\mathbb{N},
\ee
with $\Omega_i = X$ or $\psi$.
The operator ${\cal O}$ transforms according to the tensor product ${\cal V}_j^{\otimes L}$ of $L$ copies of the infinite dimensional
chiral representation ${\cal V}_j$ with superconformal spin $j=1$. The irreducible components are associated to superconformal 
primaries ${\cal O}_\alpha$ with quantum numbers $\alpha$. The lowest weight vectors $\Psi_\alpha$ in each module can be obtained by Bethe Ansatz
methods. They are eigenstates of the Cartan generators $J$, $\overline{J}$ and the quadratic Casimir $\mathbb{C}_2 = J\overline{J}$
\ba
\mathbb{C}_2\,\Psi_\alpha &=& J\,\overline{J}\,\Psi_\alpha, \\
J\,\Psi_\alpha &=& (m+L)\,\Psi_\alpha, \\
\overline J\,\Psi_\alpha &=& \overline{m}\,\Psi_\alpha.
\ea
The quantum numbers are thus $\alpha = [L, \overline{m}, m]$. 
It can be shown that $m, \overline{m}$ are non-negative integers with 
\be
1\le\overline{m}-m\le L-1.
\ee
The states associated with the highest weight at the boundary $\overline{m}-m =1$ are trivially related to the states $\mbox{Tr}(\partial_+^{\overline m} X^L)$
in the bosonic $\mathfrak{sl}(2)$ sector. Those at the opposite boundary  $\overline{m}-m =L-1$ are associated with $L$-gaugino states  
$\mbox{Tr}(\partial_+^{m} \psi^L)$. Hence, we can say that the $\mathfrak{sl}(2|1)$ sector interpolates between fully bosonic/fermionic states.
This is precisely the framework we need to prove the claimed universality.

\medskip
A nested Bethe Ansatz valid in this sector is described in \cite{Belitsky:2007zp}, according to the methods 
of~\cite{Beisert:2005fw,Belitsky:2006cp}. Up to three loops, the Bethe Ansatz equations read
\ba
\label{eq:ba}
\left(\frac{x^+_k}{x^-_k}\right)^L &=& \mathop{\prod_{j, k=1}^{\overline m}}_{j\neq k} \frac{x^-_k-x^+_j}{x^+_k-x^-_j}
\frac{1-\frac{g^2}{2\,x^+_k \, x^-_j}}{1-\frac{g^2}{2\,x^-_k \, x^+_j}}\,\cdot\,\prod_{j=1}^{\overline{m}-m-1}\frac{x^+_k-x_j^{(1)}}{x^-_k-x_j^{(1)}},\qquad k=1, \dots, \overline{m}, \\
1&=& \prod_{j=1}^{\overline m}\frac{x_k^{(1)}-x_j^+}{x_k^{(1)}-x_j^-},\qquad k= 1, \dots, \overline{m}-m-1.
\ea
In these equations, we have introduced $\overline{m}$ first level Bethe roots $\{u_k\}$ and $\overline{m}-m-1$ second level roots $\{u^{(1)}_k\}$. The notation is standard
\ba
x(u) &=& \frac{1}{2}(u+\sqrt{u^2-2\,g^2}), \\
x^\pm &=& x(u^\pm), \\
u^\pm &=& u \pm \frac{i}{2}.
\ea
The anomalous dimension is expressed in terms of the first level Bethe roots as explained in details in~\cite{Belitsky:2007zp} where a single Baxter equation for the 
first level roots is derived.

\medskip
Now to the proof. Let us consider the $L=3$ case on the gaugino boundary
\be
\overline{m}-m = L-1 = 2,\qquad\la\qquad \overline{m}-m-1=1.
\ee
Of course, it will be clear that generalizations to higher twists are possible. 
Solving the Baxter equation for even $m$, one finds that the ground state is an unpaired state with an even distribution of the first level roots
and an even Baxter function. This is similar to what happens at twist 2. There is a single second level root $u_1^{(1)}$. Let $x\equiv x(u_1^{(1)})$.
As explained in~\cite{Beisert:2005fw}, beyond one-loop, it is convenient to consider $x$ as the basic spectral parameter.
The cyclicity constraint reads
\be
\prod_{j=1}^{\overline m}\frac{x-x_j^+}{x-x_j^-}=1.
\ee
We know that the ground state is unpaired. For a non trivial even distribution of first level roots, a unique solution for $x$ is obtained if $x=0$.
This can be checked by defining the 
phase in the perturbative expansion of $x$ according to the formula
\be
x(u)\equiv\frac{u}{2}\left(1+\sqrt{1-\frac{2\,g^2}{u^2}}\right) = u-\frac{g^2}{2\,u} + \cdots.
\ee
Setting $x=0$ in the Bethe Ansatz equations Eqs.~(\ref{eq:ba}), we obtain 
\ba
\left(\frac{x^+_k}{x^-_k}\right)^{L-1} &=& \mathop{\prod_{j, k=1}^{\overline m}}_{j\neq k} \frac{x^-_k-x^+_j}{x^+_k-x^-_j}
\frac{1-\frac{g^2}{2\,x^+_k \, x^-_j}}{1-\frac{g^2}{2\,x^-_k \, x^+_j}},\qquad k=1, \dots, \overline{m}, \\
1 &=& \prod_{j=1}^{\overline m}\frac{x_j^+}{x_j^-}.
\ea
However, these are precisely the Bethe Ansatz equation for the $\mathfrak{sl}(2)\subset\mathfrak{su}(2|1)$ sector with $L-1=2$ fields and
total spin $\overline{m} = m+2$. It is well known, that these equations reproduce the correct three loop expression of $\gamma_{\rm univ}$, thus
proving our conjecture~\refeq{eq:conj}. 

\medskip
As a check, we have also computed the analytical three loop anomalous dimensions at several even spins from Eqs.~(\ref{eq:ba})
according to the methods of~\cite{Beccaria:2007cn,Kotikov:2007cy} and verified the perfect agreement with~\refeq{eq:conj},
using Eqs.~(\ref{eq:univ}).

\section{Alternative proof in the Baxter formalism}
\label{sec:baxter}

As a further analysis, we now present an alternative proof based on the analysis of the Baxter equation for the $\mathfrak{sl}(2|1)$ sector. This is 
a slightly different approach. In particular, the second level Bethe root is completely bypassed. 
The limitation is that the proposed Baxter equations admit simple polynomial Baxter functions up to $L$ loops (included) for operators with twist $L$. 
Hence, they can used to prove universality \refeq{eq:conj} between $L=2$ and $L=3$ operators at two loops only.

\medskip
The Baxter equation for the $\mathfrak{sl}(2|1)$ sector has been derived in~\cite{Belitsky:2007zp} and takes the following form 
\ba
\lefteqn{\left[\tau(x)\,\overline{\tau}(x)-(x^+\,x^-)^L\,e^{\Sigma(x^-)+\Sigma(x^+)}\right]\,Q(u) = } && \nonumber  \\
&& (x^+)^L\,e^{\Delta_+(x^+)}\,\left[\tau(x)-(x^-)^L\,e^{\Sigma(x^-)}\right]\,Q(u+i)  +  \\
&& (x^-)^L\,e^{\Delta_-(x^-)}\,\left[\overline\tau(x)-(x^+)^L\,e^{\Sigma(x^+)}\right]\,Q(u-i).  \nonumber
\ea
Here, $\tau$ and $\overline{\tau}$ are defined as 
\be
\tau(x) = (x^-)^L\,\left(1+\sum_{k\ge 1}\frac{q_k(g)}{(x^-)^L}\right), \qquad 
\overline\tau(x) = (x^+)^L\,\left(1+\sum_{k\ge 1}\frac{\overline q_k(g)}{(x^+)^L}\right),
\ee
where $q_i(g)$ are coupling dependent charges, {\em i.e.} integrals of motion. At one-loop, the only non vanishing charge (related to $\mathbb{C}_2$) is 
\be
q_2 = \overline{q}_2 = -\overline{m}(m+L).
\ee
The universal quantity $\Delta_\sigma(x)$, with $\sigma=\pm$,  admits the following three loop expansion~\cite{Belitsky:2006wg}
\ba
\Delta_\sigma(x) &=& -\frac{g^2}{x}\,\left(\log\,Q\left(\frac{i\,\sigma}{2}\right)\right)' + \\
&& -\frac{g^4}{4\,x^2}\left[
\left(\log\,Q\left(\frac{i\,\sigma}{2}\right)\right)'' + x\,
\left(\log\,Q\left(\frac{i\,\sigma}{2}\right)\right)'''\right] + {\cal O}(g^6). \nonumber
\ea
Finally, $\Sigma(x)$ is defined as the half-sum
\be
\Sigma(x) = \frac{1}{2}\left[\Delta_+(x)+\Delta_-(x)\right].
\ee

\medskip
\medskip
At twist-3 and at the gaugino boundary $\overline{m} = m+2$, we  consider the $\mathfrak{sl}(2|1)$ unpaired  ground state
with quantum numbers
\ba
\alpha &=& [L, \overline{m}, m] = [3, 2\,n+2, 2\,n],\qquad n\in \mathbb{N}, \\
q_2 &=& -\overline{m}\,(m+L) = -(2\,n+2)\,(2\,n+3)
\ea
It can be shown that, for this state,  
the polynomial Baxter function is a polynomial of degree $\overline{m}=2\,n+2$, {\em even} under $u\to -u$. Hence, at two loops level, we have simply
\be
\Sigma(x) = {\cal O}(g^4).
\ee
The Baxter equation greatly simplifies and reduces to 
\ba
\lefteqn{
\left[\tau(x)\,\overline{\tau}(x)-(x^+\,x^-)^3\right]\,Q(u) = }&& \\
&& = (x^+)^3\,e^{\Delta_+(x^+)}\,\left[\tau(x)-(x^-)^3\right]\,Q(u+i)  + (x^-)^3\,e^{\Delta_-(x^-)}\,\left[\overline\tau(x)-(x^+)^3\right]\,Q(u-i). \nonumber
\ea
The first odd charge $q_1$, as well as the higher ones $q_n$ with $n\ge 3$, vanish
\be
q_1 = \overline{q}_1 = 0,\qquad q_n = \overline{q}_n = 0, \ n\ge 3.
\ee
Imposing these conditions on $\tau$, $\overline{\tau}$ and simplifying, one obtains
\ba
\label{eq:twist2}
\lefteqn{\left[(x^+)^2+(x^-)^2+q_2\right]\,Q(u) =} &&\\
&& \qquad =  (x^+)^2\,e^{\Delta_+(x^+)}\,Q(u+i)  + (x^-)^2\,e^{\Delta_-(x^-)}\,Q(u-i). \nonumber
\ea
On the other hand, one can write down the Baxter equation for the ground state at twist-2, again at the gaugino boundary. In particular, we can 
consider the unpaired ground state with quantum numbers
\ba
\alpha' &=& [L', \overline{m}', m'] = [2, 2\,n+2, 2\,n+1],\qquad n\in \mathbb{N}, \\
q_2' &=& -\overline{m}'\,(m'+L') = -(2\,n+2)\,(2\,n+3).
\ea
This state has also a Baxter function which is an {\em even} polynomial of degree $\overline{m}' = 2\,n+2$, and  we have again $q_1=\overline{q}_1=0$ and
$q_n=\overline{q}_n=0$ for $n\ge 3$. The Baxter equation has immediately the precise form~\refeq{eq:twist2} with $q_2\to q_2'$. 
Since $q_2=q_2'$, we conclude
that the Baxter function for the state $\alpha$ is equal to the one for $\alpha'$. This leads to \refeq{eq:conj}, because the 
formula for the anomalous dimension depends only on $\alpha$ 
through $Q$~\cite{Belitsky:2007zp}.

\section{Conclusions}

In summary, we have shown that the lowest anomalous dimension of 3-gaugino operators in ${\cal N}=4$ SYM with even spin obey a remarkable universality property.
It can be expressed in terms of the universal anomalous dimension valid in the twist-2 supermultiplet. We have proved this property at three loops
as a nice exercise illustrating a general mechanism related to the hidden $\mathfrak{psu}(1|1)$ symmetries of the Bethe Ansatz equations.

\medskip
This kind of phenomena has been first discussed in~\cite{Beisert:2005fw}. Here, we have given a simple explicit example. The known two loop 
calculations are immediately reproduced, plus a novel three loop prediction. Since, supersymmetry is responsible for this 
degeneracy relating different twist operators, we believe that it would be worth to prove the universality beyond three loops 
in terms of the structure of twist-3 supermultiplets.

\medskip
We have also proved universality working in the framework of the (nested) Baxter equation, thus bypassing the analysis of the 
second level roots. However, we believe that the Bethe Ansatz equations are more enlightening, since this kind of mechanisms is known to work in 
larger sectors, like $\mathfrak{su}(1,1|2)$~\cite{Beisert:2005fw} where a Baxter equation is not yet available.

\medskip
Also, the Baxter function polynomiality breaks down at $L+1$ loop level for operators with twist $L$. This appears to be a weak point
of the Baxter approach deserving improvement. For instance, it is known that the Bethe Ansatz equations predict the correct $\gamma_{\rm univ}$
at three loops in twist-2. It seems that the Bethe Ansatz equations are more suitable to explore the degeneracies associated with the full
$\mathfrak{psu}(2,2|4)$ algebra of the ${\cal N}=4$ theory.

\acknowledgments
We thank G. Marchesini and Yu. Dokshitzer for very useful comments.

\end{document}